\documentclass[prd,aps,twocolumn,superscriptaddress,nofootinbib]{revtex4}
\usepackage[cp1251]{inputenc}
\usepackage[english]{babel}
\usepackage{amstext}
\usepackage{amssymb}
\usepackage {graphicx}
\begin{document}
 \title{INFRARED BEHAVIOUR OF GLUON AND GHOST PROPAGATORS IN QCD IN THE LANDAU GAUGE }%
 \author{P.O. FEDOSENKO}
 \affiliation{Kyiv National Taras Shevchenko University, Physics Department, Ukraine}
 \thanks{Email address: fedosenko@univ.kiev.ua}

 \begin{abstract}
Non-perturbative formalism of the generalized effective action is
used for deriving the Schwinger-Dyson equations. In order to clear
the domain of integration in the functional integral from gauge
copies a restriction to the Gribov horizon due to Zwanziger is
implemented. In this approach an asymptotic behaviour of  gluon
propagator and propagator of the Faddeev-Popov ghosts at small
momenta is studied. Such a behaviour is obtained as a result of
solving coupled Schwinger-Dyson equations in zeroth and first-order
approximation. The qualitative agreement of these results with the
ones obtained before is demonstrated and quantitative difference in
some coefficients is found.
 \end{abstract}

 \maketitle

\section{Introduction}
It is widely believed that Quantum Chromodynamics (QCD) is the
theory which describes strong interactions. The physical phenomena
at large momenta transfers are very well described by the
perturbation theory as the coupling becomes small. Asymptotic
freedom allows high energy quarks and gluons to be treated as weakly
interacting particles. This picture, however, starts to break down
at intermediate momenta and is surely inadequate at energies below a
few hundred MeV. At such scales the interaction is strong enough to
invalidate the perturbation theory and one has to employ completely
different methods to study non-perturbative phenomena like
confinement or chiral symmetry breaking. One such method is the
study of the Schwinger-Dyson equations (SDEs).

Being the non-perturbative equations of motion for Green’s
functions, SDEs plays an important role in this sector of
theoretical physics \cite{Fischer0}. Greens's functions which can be
extracted from these equations, involve an interesting physics. For
example, the non-perturbative behaviour of the infrared gluon and
ghost propagators is related to the confinement mechanism like the
Gribov-Zwanziger scenario \cite{Gribov1, Zwanziger,
Zwanziger_conf2}. Furthermore, certain consequences related to color
confinement can be obtained from the infrared behaviour of the gluon
propagator in terms of unobservable particles, the so-called
positivity violation.

Many attempts have been made to understand the gluon propagator behaviour
through SDEs. In the late seventies Mandelstam initiated the study of the
gluon SDE in the Landau gauge \cite{Mandelstam1}. Neglecting the ghost fields contribution
and imposing cancellations of certain terms in the gluon polarization tensor,
he found a highly singular gluon propagator in the infrared. This enhanced
gluon propagator was appraised for many years in the literature, firstly because
it provided a simple picture of quark confinement, since it is possible to
derive from it an inter-quark potential that rises linearly with the separation,
and secondly because the gluon propagator, which is singular as $1/q^{4}$, has enough
strength to support dynamical chiral symmetry breaking. However, these results
are discarded by simulations of QCD on the lattice \cite{lattice} where it is shown that the gluon propagator is probably infrared finite.

Later, infrared finite solutions were also found in the
Schwinger-Dyson approach \cite{Atkinson}. Considering coupled gluon
and ghost SDEs it was shown that gluon propagator is suppressed and
ghost propagator is enhanced in the infrared region.
Brief overview
of the results on the infrared properties of the gluon and ghost
propagators from SDEs existing in today's literature is made in
Section \ref{State of affairs}

In the main part of this work (Section 3) the formalism of the
generalized effective action proposed by Cornwall, Jackiw and
Tomboulis (CJT) \cite{CJT} is investigated. The effective action,
which depends on the gluon and ghost propagators and the Gribov mass
parameter is obtained. In Section~4 we construct SDEs for the gluon
and ghost propagators as well as for the Gribov parameter in
one-loop approximation. Notice, that the SDEs obtained in our work
are not exactly the same ones, discussed in literature
\cite{Zwanziger2, Lerche1, Fischer1, Fischer+}. Let us clarify this
point.

The starting point in most papers is a full system of SDEs. It is a coupled, nonlinear system
of equations, which contains dressed propagators as well as dressed vertex
functions on the right-hand side. In order to obtain a closed system of equations
it is necessary to specify suitable approximations for these equations. There are many truncation schemes, developed for this purpose, for example replacing dressed vertices by their tree-level expressions. Furthermore, in this formulation,
Gribov’s prescription of cutting off the functional integral at the Gribov horizon does not change the Schwinger-Dyson equations, but rather resolves an ambiguity in the solution of these equations.

On the other hand, in our presentation we restrict ourselves to two-loop approximation of the effective action and  obtain a closed system of SDEs without any other truncation schemes. These three equations are derived as a result of equating to zero the functional derivatives of the generalized effective action with respect to the gluon and ghost propagators and the Gribov parameter. Restriction to the (first) Gribov region appears explicitly via modification of the free gluon propagator, which enters the gluon SDE explicitly and, therefore, two other SDEs are affected via the  dressed gluon propagator too.

In Section 5 the zeroth-order analysis of the derived SDEs is carried out. It is noteworthy that at this order we get the results for the gluon propagator and horizon condition similar to those previously obtained by Gribov. It is shown that the ghost propagator obtained at this order and Gribov's one indicate the same qualitative behaviour $1/q^{4}$ in the infrared. However they differ in some coefficient.

The full analysis of the derived one-loop SDEs is developed in Section~\ref{2Loop} The infrared critical exponents are obtained as well as the numerical coefficients for the propagators. In the last Section~(7) the discussion of obtained results is presented.

\section{Nowadays state of affairs}
\label{State of affairs}
The infrared behaviour of the Landau gauge gluon and ghost propagators is an interesting
and hot subject. Up to now there are two competitive viewpoints on this problem.

Let us represent the gluon and ghost propagators, respectively, in the form:
\begin{equation}
    G^{ab}_{\mu\nu}(q) = C_{G} \frac{\widetilde{G}(q^{2})}{q^{2}}\left(g_{\mu\nu} - \frac{q_{\mu}q_{\nu}}{q^{2}}\right)\delta^{ab},
\end{equation}
\begin{equation}
    \mathcal D^{ab}(q) = C_{\mathcal D} \frac{\widetilde{\mathcal D}(q^{2})}{q^{2}} \,\delta^{ab},
\end{equation}
where $C_{G}$ and $C_{\mathcal D}$ are dimensionless constants. We will seek for the solutions of the gluon and ghost dressing functions in the form of simple power laws
\begin{equation} \label{DressedG}
    \widetilde{G}(q^{2}) \propto \left(q^{2}\right)^{-\alpha_{G}}
\end{equation}
and
\begin{equation}\label{DressedD}
  \widetilde{\mathcal D}(q^{2}) \propto \left(q^{2}\right)^{-\alpha_{\mathcal D}},
\end{equation}
respectively. Coefficients $\alpha_{G}$ and $\alpha_{\mathcal D}$ are called infrared critical exponents or anomalous dimensions.

The first point of view is as follows. Using the gluon propagator SDE as well as the ghost propagator SDE it is claimed that $\alpha_{G} + 2 \alpha_{\mathcal D} = - (4-d)/2$, \cite{Zwanziger2, Lerche1}, i.e. for the dimension of space-time $d=4$ one gets $\alpha_{G} + 2 \alpha_{\mathcal D} = 0.$ The interesting point here is that self-consistency forces an interrelation of the exponents
such that they depend on one parameter $k = \alpha_{\mathcal D} = - \alpha_{G}/2$ only. The
value of the exponent $k$ is in the range $0.5 < k \leq 1,$ \cite{Zwanziger2, Lerche1, Fischer1, Fischer+}, depending on details of the truncation of the set of SDEs.

Some of the authors report on the ambiguous result~\cite{Zwanziger2}:
\begin{equation}
\alpha_{G}\! = - 2,\,  \alpha_{\mathcal D} \!= 1 \quad
   \mbox{and} \quad
\alpha_{G}\! \approx -1.1906, \,  \alpha_{\mathcal D}\! \approx 0.5953.
\end{equation}

It is worth noticing, that the Kugo-Ojima confinement criterion
\cite{Kugo+Ojima} in terms of infrared exponents requires $k > 0,$
which means that the ghost propagator should be more singular and
the gluon less singular than a simple pole. On the other hand,
Gribov-Zwanziger confinement scenario \cite{Gribov1,
Cucchieri+Zwanziger_conf, Zwanziger_conf, Zwanziger_conf2} requires
the same condition for the ghost propagator, but it needs $k > 0.5$
for the gluon dressing function. Thus, available results for the
exponents are in good agreement with these two widespread
confinement scenarios.

However there is another point of view on this problem. Some of the
authors \cite{Non} report that $\alpha_{G} + 2 \alpha_{\mathcal D}
\neq 0$ and correct value of $\alpha_{\mathcal D}$ is close to zero
while $\alpha_{G}$ is close to 1. Thus they claim that in the
infrared limit the ghost dressing function is finite and differs
from zero.

The recent numerical results, i.e. lattice simulations, also are not
unambiguous \cite{LatticeRes}.

The aim of the present paper is independently determine the infrared
critical exponents.

\section{Cornwall-Jackiw-Tomboulis formalism} \label{CJT-form.}
We start to consider this problem with making use of the formalism of the generalized effective action proposed by J. Cornwall, R. Jackiw and E. Tomboulis \cite{CJT}. This generalization of the effective action, $\Gamma(\phi,G),$ depends not only on $\phi$ -- a possible expectation value of the quantum field $\Phi(x)$ -- but also on $G(x,y)$ -- a possible expectation value of a bilocal operator $T \Phi(x) \Phi(y)$ -- time ordered product of two fields. Physical solutions require
\begin{equation} \label{variations}
    \frac{\delta\Gamma(\phi,G)}{\delta\phi(x)} = 0, \qquad
    \frac{\delta\Gamma(\phi,G)}{\delta G(x,y)} = 0.
\end{equation}
In our derivation we identify $\Phi(x)$ with the gluon field, so $G(x,y)$ is the gluon propagator. We introduce two more quantities -- the ghost propagator $\mathcal D (x,y)$ and the so-called Gribov parameter $\gamma.$ Parameter $\gamma,$ known also as the Gribov mass, characterizes the restriction of the domain of integration in the  functional integral to the so-called Gribov horizon. This restriction is necessary due to the existence of the Gribov copies, which imply that the Landau condition, $\partial_{\mu} A_{\mu} =0$, does not uniquely fix the gauge and equivalent gauge copies still exist in the domain of integration of the functional integral \cite{Gribov1, Lections}.

We start from the generating functional for Green's functions of nonlocal, composite fields:
\begin{equation} \label{Generating Functional}
Z(J,K) = e^{\frac{i}{\hbar}W(J,K)} = \int D \Phi e^{\frac{i}{\hbar} S_{eff}},
\end{equation}
\[
   S_{eff} = \!   - \frac{1}{4}\!\!\int \!\!\! d^{4}\!x\left(F_{\mu\nu}^{a}(x)\right)^2                                                               \!\!  + \!\!\int \!\!\! d^{4}\!x\, d^{4}\!y\overline{C}^{a}\!\!(x)J^{ab}\!(x,y)C^{b}\!(y)+
\]
\begin{equation} \label{L_eff}
                + \frac{1}{2}\int \!\!\! d^{4}\!x\, d^{4}\!y A^{a}_{\mu}(x)K^{ab}_{\mu\nu}(x,y)A^{b}_{\nu}(y)
                + S_{z}(\Phi(x)) ,
\end{equation}
where
\[
S_{z}(\Phi(x))=\!\!\int \!\!\! d^{4}\!x \{
              g\gamma^{2}f^{abc} (   A^{a}_{\mu}\varphi^{bc}_{\mu}
                                   + A^{a}_{\mu}\overline{\varphi}^{bc}_{\mu} )
            - \lambda^{a}(\partial_{\mu}A^{a}_{\mu})-
\]
\[
            - \overline{C}^{a}\partial_{\mu}(D_{\mu}C)^{a}
            - \overline{\varphi}^{ac}_{\mu}\partial_{\nu}(D_{\nu}\varphi_{\mu})^{ac}
            + \overline{\omega}^{ac}_{\mu}\partial_{\nu}(D_{\nu}\omega_{\mu})^{ac}+
\]
\begin{equation} \label{L_Z}
            + g(\partial_{\nu}\overline{\omega}_{\mu}^{ac})f^{abm}(D_{\nu}C)^{b}\varphi^{mc}_{\mu}
            + 4\gamma^{4}(N^{2}-1)\}
\end{equation}
is the expression, which characterizes Zwanziger's formulation of
the Gribov horizon \cite{Zwanziger} (with $g$ being the strong
coupling constant). It is BRST-invariant together with the first
term of (\ref{L_eff}). Functional differential $D\Phi$ labels the
product of differentials of all fields entering the integrand
(effective action (\ref{L_eff})):
\begin{equation} \label{DPhi}
    D\Phi~\equiv~dA dC d\overline{C} d\lambda d\varphi d\overline{\varphi} d\omega d\overline{\omega}.
\end{equation}
Let us clarify the notations above.
Here $A$ is the gluon field with components $A^{a}_{\mu}.$
Fields $C$ and $\overline{C}$ are the Faddeev-Popov ghosts which have components $C^{a}$ and $\overline{C}^{a}.$ Multiplier $\lambda^{a}$ enforces the constraint $\partial_{\mu } A_{\mu } = 0,$ characteristic of the Landau gauge.
Fields $\varphi \equiv -\frac{1}{\sqrt{2}}(\varphi_{1} + i\varphi_{2})$ and $\overline{\varphi} \equiv                       \frac{1}{\sqrt{2}}(\varphi_{1} - i\varphi_{2})$ are the pair of boson complex fields with components
$   \varphi^{a}_{i} \equiv  \varphi^{ac}_{\mu } \equiv  f^{abc}\varphi^{b}_{\mu}$ and $f^{abc}$ are the structure constants of the $SU(N)$ gauge group.
Here we use the single index $i \equiv (\mu , c)$ for the pair of mute indices, and $i$ takes on $f=d(N^{2}-1)$
values ($d=4$ stands for the dimension of space).
The fields $\omega$ and $\overline{\omega}$ are the Grassmann fields, which have the same components as $\varphi$ and            $\overline{\varphi}.$
The vector indices $\mu,\nu$ take on $d$ values, while the color indices $a,b$ refer to the adjoint representation of the $SU(N)$ group and take on $N^{2}-1$ values.

We introduced  two bilocal sources $K^{ab}_{\mu\nu}(x,y)$ and $J^{ab}(x,y)$ for the gluon field $A^{a}_{\mu}$ and for the ghosts $\overline{C}^{a} ,C^{a} $ respectively.

Let $G^{ab}_{\mu\nu}(x,y)$ be a possible expectation value of $T A^{a}_{\mu}(x) A^{b}_{\nu}(y)$ and $\mathcal{D}^{ab}(x,y)$ stands for a possible expectation value of $T {C}^{a}(x) \overline C^{b}(y):$
\begin{equation} \label{G^ab_mu nu}
  G^{ab}_{\mu\nu}(x,y) \equiv \left\langle  T A^{a}_{\mu}(x) A^{b}_{\nu}(y)  \right\rangle,
\end{equation}
\begin{equation} \label{D^ab}
  \mathcal{D}^{ab}(x,y) \equiv \left\langle  T C^{a}(x) \overline{C}^{b}(y)  \right\rangle.
\end{equation}
From eqs. (\ref{Generating Functional}) and (\ref{L_eff}) it follows that
\begin{equation} \label{Leg Transform W-K}
   \frac{\delta W(J,K)}{\delta K^{ab}_{\mu\nu}(x,y)} = \frac{1}{2}\hbar \,G^{ab}_{\mu\nu}(x,y),
\end{equation}
\begin{equation} \label{Leg Transform W-J}
   \frac{\delta W(J,K)}{\delta J^{ab}(x,y)} = -\hbar \,\mathcal{D}^{ba}(y,x).
\end{equation}
The effective action $\Gamma(G,\mathcal{D})$ is defined as a Legendre transform of W(J,K):
\[
\Gamma(G,\mathcal{D}) = W(J,K) - \frac{\hbar}{2}\int\!\!\! d^{4}\!x\, d^{4}\!y G^{ab}_{\mu\nu}(x,y)K^{ab}_{\mu\nu}(x,y)+
\]
\begin{equation} \label{Gamma(G,D)}
 + \hbar\int \!\!\! d^{4}\!x\, d^{4}\!y \mathcal{D}^{ba}(y,x)J^{ab}(x,y).
\end{equation}
We eliminate $K$ and $J$ in favor of $G$ and $\mathcal{D}$ using the next obvious relations:
\begin{equation} \label{K}
  K^{ab}_{\mu\nu}(x,y) = -\frac{2}{\hbar}\frac{\delta \Gamma(G,\mathcal{D})}{\delta G^{ab}_{\mu\nu}(x,y)}
\end{equation}
and
\begin{equation} \label{J}
  J^{ab}(x,y) = \frac{1}{\hbar}\frac{\delta \Gamma(G,\mathcal{D})}{\delta \mathcal{D}^{ba}(y,x)}.
\end{equation}
Then according to (\ref{Gamma(G,D)}), we have
\[
 \exp {  \frac{i}{\hbar} \Gamma(G,\mathcal{D}) }
 = \exp {\frac{i}{\hbar}W(J,K)}\times
\]
\[
\times \exp \{ -\frac{i}{2}\!\int\! d^{4}\!x\, d^{4}\!y G^{ab}_{\mu\nu}(x,y)K^{ab}_{\mu\nu}(x,y)+
\]
\begin{equation} \label{exp Gamma(G,D)}
 + i\!\int\! d^{4}\!x\, d^{4}\!y \mathcal{D}^{ba}(y,x)J^{ab}(x,y)\}.
\end{equation}
In order to shorten our notations we will omit dependence on coordinates and the corresponding integration where it is obvious, keeping it in mind. Taking into account (\ref{Generating Functional}), (\ref{L_eff}) together with (\ref{K}) and (\ref{J}) we have
\[
 \exp {\frac{i}{\hbar}\Gamma(G,\mathcal{D})}
  = \!\!\!\int\!\!\! D \!\Phi
    \exp {\frac{i}{\hbar}}\{-\!\frac{1}{4}\left(F_{\mu\nu}^{a}\right)^2\!\!
           - \!\frac{1}{\hbar}A^{a}_{\mu}\frac{\delta \Gamma}{\delta G^{ab}_{\mu\nu}}A^{b}_{\nu}+
\]
\begin{equation} \label{exp Gamma}
   + \frac{1}{\hbar}\overline{C}^{a}\!\!\frac{\delta \Gamma}{\delta \mathcal{D}^{ba}}C^{b}
   + {G^{ab}_{\mu\nu} \frac{\delta \Gamma}{\delta G^{ab}_{\mu\nu}}
   + \mathcal{D}^{ba} \frac{\delta \Gamma}{\delta \mathcal{D}^{ba}}
   + S_{z}(\Phi) \}}.
\end{equation}
After straightforward computations, for the first exponent term in the last expression one gets
\[
    - \frac{1}{4}\left(F_{\mu\nu}^{a}\right)^2
     = \frac{1}{2}A^{a}_{\nu}(iD^{-1}_{F})^{ab}_{\mu\nu}A^{b}_{\mu}
     - gf^{abc}\left((\partial_{\mu}A^{a}_{\nu})A^{b}_{\mu}A^{c}_{\nu} \right)-
\]
\begin{equation} \label{first two terms}
        - \frac{1}{4}g^{2}f^{abc}f^{apr}A^{b}_{\mu}A^{c}_{\nu}A^{p}_{\mu}A^{r}_{\nu}.
\end{equation}
Here $D_{F}$ is a free gluon propagator:
\begin{equation} \label{free boson propagator}
  (iD^{-1}_{F})^{ab}_{\mu\nu} \equiv
    \left(\partial^{2}g_{\mu\nu}-\partial_{\mu}\partial_{\nu}\right)\delta^{ab}.
\end{equation}
The series expansion for $\Gamma(G,\mathcal{D})$ is
\[\Gamma(G,\!\mathcal{D}) \!\!=\!\! \frac{1}{2}i\hbar Tr \ln (G^{-1})^{ab}_{\mu\nu}\!\!\!
                    + \!\!\frac{1}{2}i\hbar Tr\!\!\left[(\widetilde{D}_{F}^{-1})^{ab}_{\mu\nu}G^{ab}_{\mu\nu}\right]\!\!
                    + \!\Gamma_{2}(G)-
\]
\begin{equation} \label{Series expansion Gamma}
                       - i\hbar Tr \ln (\mathcal D^{-1})^{ab}
                       - i\hbar Tr \!\!\left[(S_{F}^{-1})^{ab}\mathcal{D}^{ba}\right]
                       + \Gamma_{2}(\mathcal{D})
                       + \mbox{Const},
\end{equation}
where $S_{F}$ is a free ghost propagator and $\widetilde{D}_{F}$ is a free gluon propagator, modified with respect to the Zwanziger term:
\begin{equation} \label{free boson propagator feat. Zwanziger}
    \widetilde{D}_{F}^{-1} = D_{F}^{-1} + 2g^{2}\gamma^{4}N(i\partial^{2})^{-1}.
\end{equation}
The quantity $\Gamma_{2}$ is given by all two-particle irreducible vacuum graphs and it is of order $\hbar^{2}$, because the number of loops corresponds to powers of $\hbar$. The quantities
$\Gamma_{2}(G)$ and $\Gamma_{2} (\mathcal{D})$ are determined by the expansion by powers of $\hbar$:
\begin{equation} \label{Gamma_2_1(G)}
    \Gamma_{2} (G) = \hbar^{2}\Gamma^{(1)}_{2}(G) + O(\hbar^{3}),
\end{equation}
\begin{equation} \label{Gamma_2_1(D)}
 \Gamma_{2} (\mathcal{D}) = \hbar^{2}\Gamma^{(1)}_{2}(\mathcal{D}) + O(\hbar^{3}).
\end{equation}
From the expressions (\ref{exp Gamma}), (\ref{first two terms}) the series expansions (\ref{Series expansion Gamma}), (\ref{Gamma_2_1(G)}) and (\ref{Gamma_2_1(D)}), using the notation
\[
\Gamma^{'}\!(G,\!\mathcal D) \!\!\equiv \!\!
      -i\hbar Tr \ln (\mathcal{D}^{-1})^{ab}\!\! + \!\hbar^{2}\Gamma^{(1)}_{2}(\mathcal{D})\!\! -\! \hbar^{2}\mathcal{D}^{ba}\frac{\delta\Gamma^{(1)}_{2}(\mathcal{D})}{\delta \mathcal{D}^{ba}}+
\]
\vspace{-.2cm}
\begin{equation} \label{widetilde{Gamma}}
 + \frac{1}{2}i\hbar Tr \ln (G^{-1})^{ab} + \hbar^{2}\Gamma^{(1)}_{2}(G) - \hbar^{2}G^{ab}_{\mu\nu}\frac{\delta\Gamma^{(1)}_{2}(G)}{\delta G^{ab}_{\mu\nu}},
\end{equation}
one obtains apart from the constant $\frac{i}{2} \hbar Tr 1$:
\[
 \Gamma^{'}\!(G,\!\mathcal D)
 = - i\hbar\ln\! \int\!\! D \Phi \exp \frac{i}{\hbar} \{ 4\gamma^{4}(N^{2}\!-\!1)V  +
\]
\vspace{-.3cm}
\[
         + \frac{i}{2}A^{a}_{\nu}(G^{-1}\! - \!2g^{2}\gamma^{4}N(i\partial^{2})^{-1})^{ab}_{\mu\nu}A^{b}_{\mu}
+ i\overline{C}^{a}(\mathcal{D}^{-1}-
\]\vspace{-.3cm}
\[        - S^{-1}_{F})^{ab}C^{b}
         + g\gamma^{2}f^{abc}(A^{a}_{\mu}\varphi^{bc}_{\mu}+A^{a}_{\mu}\overline{\varphi}^{bc}_{\mu})
         -\! \lambda^{a}(\partial_{\mu}A^{a}_{\mu})+
\]\vspace{-.3cm}
\[
         + \overline{\Theta}_{i}\partial^{2}\Theta_{i}
         \!- gf^{abc}\!(\partial_{\mu}\!A^{a}_{\nu})A^{b}_{\mu}\!A^{c}_{\nu}
           - gf^{acb}\overline{C}^{a}\!\!\partial_{\mu}\!A_{\mu}^{c}\!C^{b}\!+
\]\vspace{-.3cm}
\[
         + gf^{acb}\overline{\omega}^{a}_{i}\!\partial_{\mu}\!A_{\mu}^{c}\!\omega^{b}_{i}
           \!- g\!f^{acb}\!\overline{\varphi}^{a}_{i}\!\partial_{\mu}\!A_{\mu}^{c}\!\varphi^{b}_{i}
           + g(\partial_{\nu}\!\overline{\omega}^{ac}_{\mu})f^{abm} \times
\]\vspace{-.3cm}
\[\times(\partial_{\nu}\!C)^{b}\!\varphi^{mc}_{\mu}
         - \frac{1}{4}g^{2}\!f^{abc}\!f^{apr}\!\!A^{b}_{\mu}\!A^{c}_{\nu}\!A^{p}_{\mu}\!A^{r}_{\nu}
           \!- \hbar A^{a}_{\mu}\!\frac{\delta\Gamma^{(1)}_{2}\!\!(G)}{\delta G^{ab}_{\mu\nu}}A^{b}_{\nu} +
\]\vspace{-.3cm}
\begin{equation} \label{Gamma Long}
        + \hbar \overline{C}^{a}\!\frac{\delta\Gamma^{(1)}_{2}\!\!(\mathcal{D})}{\delta \mathcal{D}^{ba}}C^{b}
        + g^{2}f^{abm}(\partial_{\nu}\overline{\omega}^{ac}_{\mu})A^{\lambda}_{\nu}C^{\lambda b}\varphi^{mc}_{\mu}.
\end{equation}
Here $V$ is the space-time volume of the system and we denote the sum of three terms of similar form by means of one term:
\begin{equation}\label{Theta}
 \overline{\Theta}_{i}\partial^{2}\Theta_{i} \equiv
 \left(
 -\overline{C}^{a}\partial^{2}C^{a}
 -\overline{\varphi}^{ac}_{\mu} \partial^{2} \varphi^{ac}_{\mu}
 +\overline{\omega}^{ac}_{\mu} \partial^{2} \omega^{ac}_{\mu}
 \right).
\end{equation}
Thus we have the expression (\ref{Gamma Long}) for the effective action, and we have to evaluate the functional integral. But this expression includes not only terms, quadratic on the fields, which we are able to integrate, but also cubic ones and terms of fourth order on the  fields. In order to compute it, one has to expand cubic and higher-order terms by powers of some small parameter in usual way. Following Cornwall at al., we will use $\hbar$ (to be precise, we use not $\hbar,$ but $\sqrt{\hbar}$) as such a small parameter, and this procedure is known as loop-expansion. We will retain only terms up to $\hbar = \left(\sqrt{\hbar}\right)^{2}$, this corresponds to two-loop approximation.

In order to select terms by powers $\hbar$ one has to change the scale of all 8 fields:
\begin{equation}
    \Phi \,(x)\rightarrow \sqrt{\hbar}\,\Phi \,(x).
\end{equation}
After rescaling the fields one can represent the series expansion by powers of small parameter $\hbar$ for (\ref{Gamma Long}) in such a way:
\[
\Gamma^{'}(G,\mathcal D)
= - i\hbar\ln \int D \Phi \hbar^{4} e^{\left\{ \frac{i}{\hbar}4\gamma^{4}(N^{2}-1)V + S^{q} \right\}}\times
\]
\begin{equation} \label{Gamma'}
         \times \!\! \left(\!1\! +\! \sqrt{\hbar}S^{(\sqrt{\hbar})}\! +\! \hbar S^{(\hbar)}
                  \!\!+ \!\frac{1}{2}\!\left(\!\sqrt{\hbar}S^{(\sqrt{\hbar})}\!\!\! +\! \hbar S^{(\hbar)} \right)^{2}
                  \!\!\!\!\! + \!o(\hbar)\!\!
         \right),
\end{equation}
where $S^{q}$ denotes quadratic terms on the fields:
\[
S^{q}\!\! \equiv \!\!  - \frac{1}{2}A^{a}_{\nu}\!\left(G^{\!-\!1}\!\!\!\! -\!\! 2g^{2}\!\gamma^{4}\!N(i\partial^{2})^{\!-\!1}\right)^{ab}_{\mu\nu}\!A^{b}_{\mu}\!\!
            - \!\overline{C}^{a}\!\!\!\left(\mathcal{D}^{\!-\!1}\!\!\!\! - \!\!S^{\!-\!1}_{F}\right)^{ab}\!C^{b}+
\]
\begin{equation}\label{L^quad}
            + ig\gamma^{2}\!f^{abc}(A^{a}_{\mu}\varphi^{bc}_{\mu}\!+\!A^{a}_{\mu}\overline{\varphi}^{bc}_{\mu})
              \!-\! i\lambda^{a}(\partial_{\mu}\!A^{a}_{\mu})
            \!+\! \overline{\Theta}_{i}i\partial^{2}\!\Theta_{i};
\end{equation}
The term $S^{(\sqrt{\hbar})}$ denotes the cubic ones:
\[
S^{(\sqrt{\hbar})} \equiv i(-gf^{abc}(\partial_{\mu}A^{a}_{\nu})A^{b}_{\mu}A^{c}_{\nu}
           - gf^{acb}\overline{C}^{a}\partial_{\mu}A_{\mu}^{c}C^{b}+
\]
\[
           + gf^{acb}\overline{\omega}^{a}_{i}\partial_{\mu}A_{\mu}^{c}\omega^{b}_{i}
             -gf^{acb}\overline{\varphi}^{a}_{i}\partial_{\mu}A_{\mu}^{c}\varphi^{b}_{i}+
\]
\begin{equation}
           + g(\partial_{\nu}\overline{\omega}^{ac}_{\mu})f^{abm}(\partial_{\nu}C)^{b}\varphi^{mc}_{\mu} ),
\end{equation}
and
\vspace{-.2cm}
\[
S^{(\hbar)} \equiv i(- \frac{1}{4}g^{2}f^{abc}f^{apr}A^{b}_{\mu}A^{c}_{\nu}A^{p}_{\mu}A^{r}_{\nu}
           - A^{a}_{\mu}\frac{\delta\Gamma_{2}(G)}{\delta G^{ab}_{\mu\nu}}A^{b}_{\nu} +
\]
\begin{equation}
  + \overline{C}^{a}\frac{\delta\Gamma_{2}(\mathcal{D})}{\delta \mathcal{D}^{ba}}C^{b}
    + g^{2}f^{abm}(\partial_{\nu}\overline{\omega}^{ac}_{\mu})A^{\lambda}_{\nu}C^{\lambda b}\varphi^{mc}_{\mu} )
\end{equation}
denotes the terms of the fourth order on the  fields.

Carrying the term with $\gamma$ in front of the sign of the integral in eq. (\ref{Gamma'}), then collecting terms by powers of $\hbar$ in the last multiplier of this equation and retaining only terms of order $\hbar$, we obtain the  two-loop contribution to the effective action of the theory~:
\[
\Gamma^{'}(G,\mathcal D)
    =  4\gamma^{4}(N^{2}-1)V-
\]
\begin{equation} \label{widetilde{Gamma}with hbar}
  - i\hbar\ln \int D \Phi \hbar^{4} e^{S^{q}}
         \left[1 + \hbar\left(S^{(\hbar)} + \frac{1}{2} (S^{(\sqrt{\hbar})})^{2}\right)\right].
\end{equation}
Expanding in $\hbar$ we obtain
\[
\Gamma^{'}(G,\mathcal D) - 4\gamma^{4}(N^{2}-1)V=
\]
\begin{equation} \label{two terms}
    = - i\hbar \ln \int D \Phi \hbar^{4}e^{S^{q}}
       - i\hbar^{2}  \left\langle S^{(\hbar)} + \frac{1}{2} (S^{(\sqrt{\hbar})})^{2} \right\rangle,
\end{equation}
The first term is the result of integration $e^{S^{q}}$ on all 8 fields (see the notation (\ref{DPhi})), together with  the notations (\ref{L^quad}) and (\ref{Theta}). It gives, omitting unimportant constants $-i\hbar \ln\hbar^{4}$ and $\frac{i}{2}\hbar Tr \ln \left[\partial_{\mu} \partial_{\nu} \right]$:
\begin{equation} \label{first term of two}
  - i\hbar \ln \int D \Phi \hbar^{4}e^{S^{q}}
=   - i\hbar \ln \left[Det(\mathcal{D}^{-1})^{ab}\right].
\end{equation}
Note that integration over the fields $\varphi, \overline{\varphi}$ and $\omega, \overline{\omega}$ gives unity as well as over $A$ and $\lambda.$

The second term of (\ref{two terms}) is a sum of all vacuum expectation values of all the field configurations, included inside brackets $\left\langle \, \right\rangle.$ One has to write down all these averages explicitly, making possible use of the Wick theorem for the computing every vacuum expectation value.
Thus, all configurations with non-zero expectation values, which contribute to the second term of eq. (\ref{two terms}) are written as follows:
\[
    \left\langle S^{(\hbar)} + \frac{1}{2} (S^{(\sqrt{\hbar})})^{2} \right\rangle
 =  -  \frac{i}{4}g^{2}f^{abc}f^{apr}
       \left\langle
          A^{b}_{\mu}A^{c}_{\nu}A^{p}_{\mu}A^{r}_{\nu}
       \right\rangle-
\]
\[
    -i\frac{\delta\Gamma_{2}(G)}{\delta G^{ab}_{\mu\nu}}
       \left\langle
        A^{a}_{\mu}A^{b}_{\nu}
       \right\rangle
    +i\frac{\delta\Gamma_{2}(\mathcal D)}{\delta \mathcal D^{ba}}
       \left\langle
        \overline{C}^{a}C^{b}
       \right\rangle -
\]
\[
     - \frac{1}{2}g^{2}f^{abc}f^{prs}
       \left\langle
          (\partial_{\mu}A^{a}_{\nu})A^{b}_{\mu}A^{c}_{\nu} (\partial_{\rho}A^{p}_{\sigma})A^{r}_{\rho}A^{s}_{\sigma}
       \right\rangle-
\]
\begin{equation} \label{sum of averages}
    - \frac{1}{2}g^{2}f^{acb}f^{prs}
     \left\langle
  \overline{\Theta}^{a}(\partial_{\mu}A_{\mu}^{c}\Theta^{b})\overline{\Theta}^{p}(\partial_{\nu}A_{\nu}^{r}\Theta^{s})
     \right\rangle .
\end{equation}
Let us consider these expectation values, making use of the Wick theorem. Here we write down only the result of the computations.
\[
    -\!\frac{1}{2}g^{2}\!f^{abc}\!f^{prs} \!\!
       \left\langle \!
          \partial_{\mu}^{(x)}\!\!A^{a}_{\nu}\!(x)A^{b}_{\mu}\!(x)A^{c}_{\nu}\!(x) \partial_{\rho}^{(y)}\!\!A^{p}_{\sigma}\!(y)A^{r}_{\rho}\!(y)A^{s}_{\sigma}\!(y)\!
       \right\rangle\!  =
\]
\[
 = \!\!\frac{1}{2}g^{2}(\!f^{abc})^{2}\!
            \{ \!- G_{\nu\rho}(x,y)\!\!\left(\partial^{(x)}_{\mu}G_{\nu\sigma}(x,y)\!\right) \!\!                                                                      \left(\partial^{(y)}_{\rho}G_{\mu\sigma}(x,y)\right)\!+
\]
\begin{equation} \label{average for Gluon at x,y}
              + \,G_{\mu\rho}(x,y)\left(\partial^{(x)}_{\mu}G_{\nu\sigma}(x,y)\right)                                                                       \left(\partial^{(y)}_{\rho}G_{\nu\sigma}(x,y)\right)
            \}.
\end{equation}
For four gluon fields at the same space-time point $A^{a}_{\mu}(x),$ one gets:
\[
 \frac{i}{4}g^{2}f^{abc}f^{apr}
       \left\langle  A^{b}_{\mu}(x)A^{c}_{\nu}(x)A^{p}_{\mu}(x)A^{r}_{\nu}(x) \right\rangle
 =\frac{i}{4}g^{2}(f^{abc})^{2}\times
\]
\begin{equation}
 \times  \{G_{\mu\nu}(x,x)G_{\mu\nu}(x,x)- G_{\mu\mu}(x,x)G_{\nu\nu}(x,x) \}.
\end{equation}
For anti-commuting ghosts (the first three fields are located in the same space-time point, say $x$, and let $y$ be space-time point for the location of the last three fields), one gets:
\[
 - \frac{1}{2}g^{2}f^{acb}f^{prs}
        \left\langle
      (\partial_{\mu}^{(x)}\overline{C}^{a})A_{\mu}^{c}C^{b} |_{(x)}
      (\partial_{\nu}^{(y)}\overline{C}^{p})A_{\nu}^{r}C^{s} |_{(y)}
        \right\rangle =
\]
\begin{equation}
    = - \frac{1}{2}g^{2}(f^{acb})^{2}
      \{
      \partial_{\mu}^{(x)}\mathcal D(y,x) \partial_{\nu}^{(y)}\mathcal D(x,y) G_{\mu\nu}(x,y)
      \} .
\end{equation}
Now with a close analogy to the ghosts, for Grassmann fields $\omega,\overline{\omega} $ we get
\[
 \frac{1}{2}g^{2}f^{acb}f^{prs}
        \left\langle
      (\partial_{\mu}^{(x)}\overline{\omega}^{a}_{i}) A_{\mu}^{c}\omega^{b}_{i} |_{(x)}
      (\partial_{\nu}^{(y)}\overline{\omega}^{p}_{j}) A_{\nu}^{r}\omega^{s}_{j} |_{(y)}
        \right\rangle=
\]
\begin{equation}
    = \frac{1}{2}g^{2}(f^{acb})^{2}
      \{
      \partial_{\mu}^{(x)}G^{\omega}_{ji}(y,x) \partial_{\nu}^{(y)}G^{\omega}_{ij}(x,y) G_{\mu\nu}(x,y)
      \}.
\end{equation}
Here $G^{\omega}_{ij}(x,y)$ is a tree level propagator for the fields $\omega, \overline{\omega}$ and indices $i,j$ take on $d(N^{2}-1)$ values.
For a pair of complex conjugate bosonic fields $\varphi, \overline{\varphi}$ we get
\[
 \frac{1}{2}g^{2}f^{acb}f^{prs}
        \left\langle
      (\partial_{\mu}^{(x)}\overline{\varphi}^{a}_{i}) A_{\mu}^{c}\varphi^{b}_{i} |_{(x)}
      (\partial_{\nu}^{(y)}\overline{\varphi}^{p}_{j}) A_{\nu}^{r}\varphi^{s}_{j} |_{(y)}
        \right\rangle =
\]
\begin{equation} \label{average for varphi}
    = - \frac{1}{2}g^{2}(f^{acb})^{2}
      \{
      \partial_{\mu}^{(x)}G^{\varphi}_{ij}(x,y) \partial_{\nu}^{(y)}G^{\varphi}_{ij}(x,y) G_{\mu\nu}(x,y)
      \},
\end{equation}
where $G^{\varphi}_{ij}(x,y)$ is a tree level propagator for the fields $\varphi, \overline{\varphi}$ and indices $i,j$ are the same ones as in the previous case.

Finally, for the quantity $\Gamma^{'}(G,\mathcal D)$ from expressions (\ref{two terms}) - (\ref{average for varphi}) one has:
\[\Gamma^{'}(G,\mathcal D)
 =  \frac{i}{2} \hbar Tr 1
   + 4\gamma^{4}(N^{2}-1)V
   - i\hbar Tr \ln (\mathcal{D}^{-1})^{ab}-
\]
\[
  - \hbar^{2}
          \{
              G^{ab}_{\mu\nu} \frac{\delta\Gamma^{(1)}_{2}(G)}{\delta G^{ab}_{\mu\nu}}
            + \mathcal{D}^{ba}\frac{\delta\Gamma^{(1)}_{2}(\mathcal D)}{\delta \mathcal{D}^{ba} }
            + \frac{i}{4}g^{2}(f^{abc})^{2}\times
\]
\[
 \times (G_{\mu\nu}G_{\mu\nu}\!\!- \!G_{\mu\mu}G_{\nu\nu})\!
 +\! \frac{i}{2}g^{2}(f^{acb})^{2} [
            - \partial_{\mu}^{(x)}\mathcal D \partial_{\nu}^{(y)}\mathcal D G_{\mu\nu}-
\]
\[
            - \partial_{\mu}^{(x)}\!G^{\omega}_{ji} \partial_{\nu}^{(y)}G^{\omega}_{ij}G_{\mu\nu}
           \!\! +\! \partial_{\mu}^{(x)}\!G^{\varphi}_{ij} \partial_{\nu}^{(y)}\!G^{\varphi}_{ij}G_{\mu\nu} ]
            \!+ \!\frac{i}{2}g^{2}(f^{abc})^{2} \times
\]
\begin{equation} \label{Gamma'_1}
            \times \left[- G_{\nu\rho}(\partial_{\mu}G_{\nu\sigma}) (\partial_{\rho}G_{\mu\sigma})
            + G_{\mu\rho}(\partial_{\mu}G_{\nu\sigma}) (\partial_{\rho}G_{\nu\sigma}) \right]
          \}.
\end{equation}
As it follows from the explicit definition of the propagators $G^{\omega}_{ij}$ and $G^{\varphi}_{ij}$ (it can be found during the explicit integration of the expression (\ref{first term of two}) for $S^{q}$ with the notations (\ref{L^quad}) and (\ref{Theta})),
$
  G^{\omega}_{ij} \equiv G^{\varphi}_{ij}.
$
Inasmuch as these propagators enter  the eq. (\ref{Gamma'_1}) via terms of equivalent form, their whole contribution cancels:
\begin{equation}
 -  \partial_{\mu}^{(x)}G^{\omega}_{ij}\partial_{\nu}^{(y)} G^{\omega}_{ij}G_{\mu\nu}
 +  \partial_{\mu}^{(x)}G^{\varphi}_{ij}\partial_{\nu}^{(y)}G^{\varphi}_{ij}G_{\mu\nu}
 = 0.
\end{equation}
Thus, the effective action has no dependence on auxiliary unphysical fields $\varphi, \overline{\varphi}$ and $\omega, \overline{\omega}.$

Finally, taking into account that for the gauge group $SU(N)$:
\begin{equation}
    f^{ade} f^{bde}= N \delta^{ab},
\end{equation}
the explicit expression (\ref{Series expansion Gamma}) for the effective action of the theory reads as:
\[  \Gamma(G,\mathcal{D}) =  4 \gamma^{4}(N^{2}\!\!-\!1)V
                            + \frac{1}{2}i\hbar Tr
                               [   \,\, \ln (G^{-1})^{ab}_{\mu\nu} +
\]
\[                    + (\widetilde{D}_{F}^{-1})^{ab}_{\mu\nu}
                       G^{ab}_{\mu\nu}\! - \!1]
                          \!- \!i\hbar Tr
                             \left[
                               \ln (\mathcal D^{-1})^{ab}
                               \!\!+\!( S_{F}^{-1})^{ab}\mathcal{D}^{ba}
                               \!-\!1
                             \right]
                          -
\]
\[                        - \frac{1}{2}i\hbar^{2}g^{2}N \delta^{cc}
                             \{
                             \frac{1}{2} (G_{\mu\nu}G_{\mu\nu}\!\!-\! G_{\mu\mu}G_{\nu\nu})
                         \! -  \partial_{\mu}^{(x)}\mathcal D \partial_{\nu}^{(y)}\mathcal D G_{\mu\nu}-
\]
\begin{equation} \label{Gamma in coordinate space}
                            - G_{\nu\rho}(\partial_{\mu}G_{\nu\sigma}) (\partial_{\rho}G_{\mu\sigma})
                          + G_{\mu\rho}(\partial_{\mu}G_{\nu\sigma}) (\partial_{\rho}G_{\nu\sigma})
                              \}.
\end{equation}
It is convenient to rewrite the expression (\ref{Gamma in coordinate space}) in momentum space, using the Fourier-transformed propagators, defined as follows:
\begin{equation}
    G(p) = \int d^{4}x e^{\frac{i}{\hbar}p(x-y)}G(x-y);
\end{equation}
(Inasmuch as we seek for the ground state, we consider the case of translation-invariant solutions, i.e., we take $G(x,y)$ to be a function only of $x-y$, and similarly for the other propagators.)

Passing to momentum space, for the eq. (\ref{Gamma in coordinate space}), divided by the volume $V$, we get
\[
    \frac{1}{V}\widetilde{\Gamma}(G,\mathcal{D})
    = 4\gamma^{4}(N^{2}-1) +
\]
\[
    + \frac{1}{2}i\hbar \!\!\!\int\!\! \frac{d^{4}p}{(2\pi\hbar)^{4}}
       \left(
               Tr \ln G^{-1}(p)
             + Tr \widetilde{D}^{-1}_{F}(p)G(p) -1
       \right) -
\]
\[
      - i\hbar \!\!\!\int\!\! \frac{d^{4}p}{(2\pi\hbar)^{4}}
       \left(
               Tr \ln \mathcal D^{-1}(p)
             + Tr S^{-1}_{F}(p)\mathcal D(p) -1
       \right) -
\]
\[
 - \frac{i}{4}\hbar^{2}\!g^{2}\!N\delta^{cc}
        \!\!\!\int\!\!\! \frac{d^{4}\!p\, d^{4}\!r}{(2\pi\hbar)^{8}} \{ G_{\mu\nu}(p)G_{\mu\nu}(r)\! - \!G_{\mu\mu}(p)G_{\nu\nu}(r)\} -
\]
\[
 - \frac{i}{2}g^{2}N\delta^{cc} \!\!\! \int\!\!\frac{d^{4}\!p\, d^{4}\!r}{(2\pi\hbar)^{8}} \{
   \mathcal D(p)\mathcal D(r)G_{\mu\nu}(p-r)p_{\mu}r_{\nu} +
\]
\[
     + G_{\nu\rho}(p)G_{\nu\sigma}(p+r)G_{\mu\sigma}(r)(p+r)_{\mu}r_{\rho} -
\]
\begin{equation} \label{Gamma in momentum space}
     - G_{\mu\rho}(p)G_{\nu\sigma}(p+r)G_{\nu\sigma}(r)(p+r)_{\mu}r_{\rho}\}.
\end{equation}
Thus we have the two-loop generalized effective action, written in the momentum representation.

\section{Schwinger-Dyson equations}\label{SDEs}
In order to obtain a closed system of equations on the gluon propagator $G_{\mu\nu}^{ab}(p)$, the ghost propagator $\mathcal D^{ab}(p)$ and the Gribov parameter $\gamma$ one has to take variational derivatives of the effective action (\ref{Gamma in momentum space}) with respect to these parameters and equate each of them to zero (see eq. (\ref{variations})).

Before doing this, it is worth to remember that in the momentum representation the expression (\ref{free boson propagator feat. Zwanziger}) for the free boson propagator modified with respect to the Zwanziger term is given by
\begin{equation} \label{free boson propagator feat. Zwanziger in p-space}
    \widetilde{D}^{-1}_{F}(p)
     = D^{-1}_{F}(p) + 2g^{2}\gamma^{4}N \frac{i\hbar^{2}}{p^{2}}.
\end{equation}
From here on we will put $\hbar=1.$

Varying expression (\ref{Gamma in momentum space}) with respect to $\gamma$, we get
\[
 \frac{1}{V}\frac{\delta\widetilde{\Gamma}(G,\mathcal{D})}{\delta \gamma}
 = 0 =
\]

\begin{equation} \label{Gamma/gamma}
 = 4\gamma^{3}\left[4(N^{2}-1)
   -g^{2}N \int \frac{d^{4}p}{(2\pi)^{4}} Tr \frac{G^{ab}_{\mu\nu}(p)}{p^{2}}\right],
\end{equation}
and after rejecting the trivial solution $\gamma = 0$, it looks as
\begin{equation} \label{Gor.Cond.01}
    1 = \frac{N g^{2}}{4(N^{2}-1)}\int \frac{d^{4}p}{(2\pi)^{4}} Tr \frac{G^{ab}_{\mu\nu}(p)}{p^{2}}.
\end{equation}
Variation with respect to $\mathcal{D}^{ab}(q)$ is given by
\[
 \frac{1}{V}\frac{\delta\widetilde{\Gamma}(G,\mathcal{D})}{\delta \mathcal D^{ab}(q)}
 = 0
 = i \left(\mathcal D^{-1}(q) - S^{-1}_{F}(q)\right)^{ab}+
\]
\[
   - \frac{i}{2}g^{2}N \delta^{ab}
    [\int \frac{d^{4}r}{(2\pi)^{4}}
         \mathcal D(r)G_{\mu\nu}(q-r)q_{\mu}r_{\nu} +
\]
\begin{equation}
       +      \int \frac{d^{4}p}{(2\pi)^{4}}\mathcal D(p) G_{\mu\nu}(p-q)p_{\mu}q_{\nu}].
\end{equation}
Due to the property $G(r) = G(-r)$ this expression after changing of mute indices $p\rightarrow r$ and $\mu\rightarrow \nu$ in the last term, can be written in the form
\[
    \left(\mathcal D^{-1}(q)\right)^{ab}
    = \left(S^{-1}_{F}(q)\right)^{ab}+
\]
\begin{equation} \label{Ghost propagator Eq.}
     + N g^{2}\delta^{ab}\int \frac{d^{4}r}{(2\pi)^{4}} \mathcal D(r) G_{\mu\nu}(q-r)q_{\mu}r_{\nu}.
\end{equation}
Variation with respect to $G^{ab}_{\mu\nu}(q)$ is given by
\[
  \frac{1}{V}\frac{\delta\widetilde{\Gamma}(G,\mathcal{D})}{\delta G^{ab}_{\mu\nu}(q)}
 =  0 = \frac{i}{2}\! \left(- G^{-1}(q)\! +\! \widetilde{D}^{-1}_{F}(q)\right)^{ab}_{\mu\nu} \!\!\!
 -  \frac{i}{2}g^{2}N\delta^{ab} \times
\]
\begin{equation} \label{Gluon propagator Eq.}
       \times \!\!\!\int\frac{d^{4}p}{(2\pi)^{4}}\mathcal D(p)\mathcal D(p-q) p_{\mu}(p-q)_{\nu}
        + (\mbox{gluon loops}).
\end{equation}
Some simplifications can be done in the expressions above. In the eq. (\ref{Gor.Cond.01}) we take the trace on color and Lorentz indices. In the Landau gauge the gluon propagator is transverse on vector indices:
\begin{equation} \label{Transversality}
    G^{ab}_{\mu\nu}(p) = G(p^{2})\mathcal P_{\mu\nu}(p)\delta^{ab},
\end{equation}
where
\begin{equation}
 \mathcal P_{\mu\nu}(p)\equiv \left[g_{\mu\nu} - \frac{p_{\mu}p_{\nu}}{p^{2}}\right]
\end{equation}
is the transverse projector.
In the eq. (\ref{Ghost propagator Eq.}) we shift the integration variable according to $r\rightarrow r+q$ and use eq.~ (\ref{Transversality}). The similar shift of integration variable will be applied to the eq. (\ref{Gluon propagator Eq.}).

Now we are ready to write down a closed system of equations of motion for the Gribov parameter, the ghost and gluon propagators:

\begin{equation} \label{Gorizon condition full}
 1 = \frac{3}{4}N g^{2} \int \frac{d^{4}p}{(2\pi)^{4}} \frac{G(p)}{p^{2}};
\end{equation}
\[
 \left(\mathcal D^{-1}(q)\right)^{ab}
    = \left(S^{-1}_{F}(q)\right)^{ab} +
\]
\begin{equation} \label{Equation on Ghost propagator}
     + N g^{2}\delta^{ab}\int \frac{d^{4}r}{(2\pi)^{4}} \mathcal D(q+r) G(r^{2})                                                                          \left[\frac{r^{2}q^{2} - (qr)^{2}}{r^{2}}\right];
\end{equation}
\[
 \left(G^{-1}(q)\right)^{ab}_{\mu\nu}
 =  \left(\widetilde{D}^{-1}_{F}(q)\right)^{ab}_{\mu\nu}
 -g^{2}N\delta^{ab} \times
\]
\begin{equation} \label{Equation on Gluon propagator}
 \times \!\!\!\int\!\!\!\frac{d^{4}p}{(2\pi)^{4}}
\{ \mathcal D(p+q)\mathcal D(p) (p+q)_{\mu}p_{\nu} \}
+ (\mbox{gluon loops}).
\end{equation}
It is worth noticing, that the Gribov parameter $\gamma$ does not enter its equation of motion (\ref{Gorizon condition full}) explicitly, but through the modified free gluon propagator $\widetilde{D}_{F}(q)$ (according to (\ref{free boson propagator feat. Zwanziger in p-space})), which is present in equation (\ref{Equation on Gluon propagator}) for the gluon propagator.

We wish to determine the asymptotic form of the propagators at low momentum. For this purpose we let the external momentum in the SDEs written above be asymptotically small. In this case the loop integration will be dominated  by asymptotically small loop momentum, so the propagators inside the integrals may also be replaced by their asymptotic values. As one can see from Section 2, where it is argued that the ghost propagator should be more singular and the gluon less singular than a simple pole, gluon loops integration at low momentum in eq. (\ref{Equation on Gluon propagator}) may be neglected.

Let us consider the eq. (\ref{Equation on Ghost propagator}). One of the form of the horizon conditions, that guaranties the absence of the Gribov copies, is written as \cite{Zwanziger_Hor.Cond.}
\[
 \lim_{q \rightarrow 0} \left[q^{2}\mathcal D(q^{2})\right]^{-1} = 0.
\]
To impose this condition, we divide eq. (\ref{Equation on Ghost propagator}) by $q^{2},$ and obtain (after the factorization of color indices)
\begin{equation}
0 = i + N g^{2}\int \frac{d^{4}r}{(2\pi)^{4}} \mathcal D(r^{2}) G(r^{2})                                                                          \left[1 - \cos(\hat{qr})\right].
\end{equation}
Subtracting this equation from the previous one (\ref{Equation on Ghost propagator}), we get
\[
  \mathcal D^{-1} (q^{2})
    = - N g^{2}\int \frac{d^{4}r}{(2\pi)^{4}} G(r^{2})[\mathcal D(r^{2})-
\]
\begin{equation}
 - \mathcal D((q+r)^{2})] \left[\frac{r^{2}q^{2} - (qr)^{2}}{r^{2}}\right].
\end{equation}
We now turn to the SDE for the gluon propagator~(\ref{Equation on Gluon propagator}) with comments below it. We apply the transverse projector and take the trace on color and Lorentz indices and obtain taking into account eq. (\ref{free boson propagator feat. Zwanziger in p-space}):
\[
G^{-1}(q^{2})
 = iq^{2} + 2Ng^{2}\gamma^{4}\frac{i}{q^{2}}-
\]
\begin{equation}
  - \frac{Ng^{2}}{3} \int\frac{d^{4}p}{(2\pi)^{4}}
   \mathcal D((p+q)^{2})\mathcal D(p^{2}) \left[ \frac{p^{2}q^{2} -(pq)^{2}}{q^{2}}\right].
\end{equation}
Notice, that in the limit $q\rightarrow 0$ the first term vanishes.

In order to simplify further analysis and compare our results with other authors, we rewrite SDEs for the gluon and ghost propagators in Euclidean space:

\[
  \mathcal D^{-1} (q^{2})
    = N g^{2}\!\!\!\int\!\!\! \frac{d^{4}p}{(2\pi)^{4}} G(p^{2})\left[\mathcal D(p^{2}) - \mathcal D((q+p)^{2})\right] \times
\]
\begin{equation} \label{Euclidean Ghost SDE}
  \times  \left[\frac{p^{2}q^{2} - (pq)^{2}}{p^{2}}\right].
\end{equation}

\[
 G^{-1}(q^{2})
 = 2g^{2}\gamma^{4}N\frac{1}{q^{2}}
  + \frac{Ng^{2}}{3}\!\!\! \int\!\!\!\frac{d^{4}p}{(2\pi)^{4}}
   \mathcal D((p+q)^{2})\mathcal D(p^{2}) \times
\]
\begin{equation} \label{Euclidean Gluon SDE}
 \times \left[ \frac{p^{2}q^{2} -(pq)^{2}}{q^{2}}\right].
\end{equation}
This system of SDEs for the gluon and ghost propagators in Landau gauge, written in Euclidean space, will be analyzed in next sections.

\section{Zeroth-order analysis of SDEs} \label{1Loop}
Now we consider the first approximation which corresponds to retaining only the first term in the right-hand side of (\ref{Equation on Gluon propagator}) without any changes in other equations. It is in fact the zeroth-order approximation. In eq. (\ref{Equation on Gluon propagator}) the contribution at this order gives
\begin{equation} \label{G(1)}
    \left(G^{(1)}(q)\right)^{ab}_{\mu\nu}
     =  \frac{q^{2}}{q^{4} + 2Ng^{2}\gamma^{4}}\delta^{ab}\left(g_{\mu\nu} - \frac{q_{\mu}q_{\nu}}{q^{2}}\right).
\end{equation}
To see this one has to inverse left-hand side and the first term in the right-hand side of eq. (\ref{Equation on Gluon propagator}) and rewrite the result in Euclidean space.

After substitution of this approximation to the Euclidean form of eq. (\ref{Gorizon condition full}), we get
\begin{equation} \label{Gap equation for gamma}
    1   = \frac{3}{4} Ng^{2} \int \frac{d^{4}q}{(2\pi)^{4}}
         \frac{1}{q^{4} + 2g^{2}N\gamma^{4}}.
\end{equation}
Note, that the gluon propagator (\ref{G(1)}) and the gap equation (\ref{Gap equation for gamma}) were first derived by Gribov \cite{Gribov1}. Furthermore, so-called horizon condition (\ref{Gap equation for gamma}) was obtained by making use of the zeroth-order approximation (\ref{G(1)}), so we see that SDE for the Gribov parameter (\ref{Gorizon condition full}) defines full horizon condition.

We now turn to the ghost SDE and apply zeroth-order approximation (\ref{G(1)}) to it. For this purpose it is more convenient to use Euclidean form of eq. (\ref{Ghost propagator Eq.}). Substituting the explicit expression for $G_{\mu\nu}(q-r)$ (according to (\ref{G(1)})) in the horizon condition (\ref{Gap equation for gamma}) one gets
\vspace{1cm}
\[
 \mathcal D^{-1}(q)
 = Ng^{2} q_{\mu}q_{\nu} \int \frac{d^{4}r}{(2\pi)^{4}}
      \frac{1}{r^{4} + \lambda^{4}}\left(g_{\mu\nu} - \frac{r_{\mu}r_{\nu}}{r^{2}}\right)\times
\]
\begin{equation} \label{Int. Eq.}
      \times\left[1 - r^{2}\mathcal D(q-r) \right],
\end{equation}
with $\lambda^{4} \equiv 2Ng^{2}\gamma^{4}.$

This is a nonlinear integral equation for the $\mathcal D$~-~function and we now have to solve it. One of the possible ways to do this is iteration scheme with a  free ghost propagator $S_{F}(q)= 1/q^{2}$ as a zeroth-order approximation $\mathcal D_{(0)}(q)$ of $\mathcal D(q)$. In other words, in order to obtain first-order approximation $\mathcal D_{(1)}^{-1}(q)$ of $\mathcal D^{-1}(q)$, we have to replace $\mathcal D(q-r) \rightarrow \mathcal D_{(0)}(q-r)\equiv 1/(q-r)^{2}$ in the right-hand side of eq. (\ref{Int. Eq.}). Doing this we get
\begin{equation} \label{expr.for Ghosts}
 \mathcal D_{(1)}^{\!-1}\!(q)
  =\! Ng^{2}\! q_{\mu}q_{\nu} \!\!\!\int\!\!\! \frac{d^{4}r}{(2\pi)^{4}}
       \frac{1}{r^{4}\! +\! \lambda^{4}}\left(g_{\mu\nu} \!-\! \frac{r_{\mu}r_{\nu}}{r^{2}}\right)\!\!\frac{q^{2}\!\! -\!\! 2qr}{(q\!-\!r)^{2}}.
\end{equation}

Notice, that exactly the same expression was used by Gribov in order to obtain the infrared behaviour of the ghost propagator \cite{Gribov1}.

We emphasize that while Gribov in fact solved the equation for the ghost propagator approximately, we derive some integral equation and we look for the exact solutions to it.

After integration of the right-hand side of (\ref{expr.for Ghosts}) for the first-order approximation of $\mathcal D(q)$ one gets \cite{Lections}:
\begin{equation} \label{Ghost Pr.}
 \mathcal D_{(1)}(q)
  =  \frac{128 \pi \lambda^{2}}{3 g^{2}N} \frac{1}{q^{4}}.
\end{equation}
Notice, that further making use of such iterational scheme for the eq. (\ref{Int. Eq.}) results in incompatible system of equations. Thus the natural question appears: is the iterational procedure, used for solving the equation (\ref{Int. Eq.}), admissible? The obvious answer is "no". Because this iteration procedure is not converging.  Consequently, perturbation theory does not work and the result, obtained above, is in general incorrect, though it gives qualitatively right behaviour.
In order to solve the equation (\ref{Int. Eq.}) correctly some another method is needed.

Let us try to solve the equation for the ghost propagator (\ref{Int. Eq.}), written in the form, similar to (\ref{Euclidean Ghost SDE}) (with the explicit gluon propagator (\ref{G(1)})) in the infrared:
\[
 \left[g\mathcal D(q^{2})\right]^{-1} = N g \int \frac{d^{4}r}{(2\pi)^{4}}
     \left[\mathcal D(r^{2}) - \mathcal D\left((q+r)^{2}\right)\right] \times
\]
\begin{equation}
     \times\frac{\left[r^{2}q^{2} - (q r)^{2}\right]}{r^{4}+\lambda^{4}}.
\end{equation}
In order to solve this equation, one can seek for the solution in the form
\begin{equation}
    g\mathcal D(q^{2}) = \frac{A}{\left(q^{2}\right)^{1+\alpha}}.
\end{equation}
Here $A$ is some coefficient and $\alpha$ is called infrared critical exponent. After evaluation of the integrals, one gets
\begin{equation}
    A^{2} = \frac{32 \pi^{2} \lambda^{4}}{N}.
\end{equation}
Thus, the exact solution of the equation (\ref{Int. Eq.}), which  describes the infrared behaviour of the ghost propagator is
\begin{equation} \label{mathcal D(1)}
    \mathcal D(q^{2}) = 4\sqrt{2}\frac{\pi\lambda^{2}}{g \sqrt{N}} \frac{1}{q^{4}}.
\end{equation}
Notice, that the previously obtained solution (\ref{Ghost Pr.}) qualitatively coincides with ours, but has another coefficient. Thus our zeroth-order analysis qualitatively confirmed that in the infrared region the gluon propagator is suppressed (\ref{G(1)}), while the ghost propagator is enhanced~(\ref{mathcal D(1)}).

\section{Full analysis of the derived one-loop SDEs} \label{2Loop}
We now turn to the SDEs (\ref{Euclidean Ghost SDE}) and (\ref{Euclidean Gluon SDE}). One can solve these equations using  "weak-angle-dependence"\ approximation for the ghosts:
\begin{equation} \label{weak angle approx.}
    \mathcal D\left((q+p)^{2}\right) \approx \mathcal D(q^{2}) \Theta(q^{2}-p^{2}) + \mathcal D(p^{2}) \Theta(p^{2}-q^{2}).
\end{equation}
Here $\Theta (x)$ is the ordinary step function.
Substituting this approximate expression to the ghost SDE (\ref{Euclidean Ghost SDE}) and gluon SDE (\ref{Euclidean Gluon SDE}) and then carrying out the angles integration, one gets
\begin{equation}\label{Gh to deff}
 \mathcal D^{-1}(q^{2})
= A q^{2}
    \int_{0}^{q^{2}} dp^{2}\, p^{2} G(p^{2})\left[\mathcal D(p^{2})-\mathcal D(q^{2})\right],
\end{equation}
\begin{equation} \label{to diff}
    G^{-1}(q^{2})
  \!\!= \!\frac{\lambda^{4}}{q^{2}}
  + B \left[ \!\!\int_{0}^{q^{2}}\!\!\!\!\!\!\! dp^{2} p^{4} \mathcal D(p^{2}) \mathcal D(q^{2})
           \!\!+ \!\!\!\int_{q^{2}}^{\infty}\!\!\!\!\!\!\! dp^{2} p^{4} \mathcal D^{2}\!(\!p^{2}\!)\right],
\end{equation}
with
\begin{equation} \label {A}
    A \equiv \frac{3Ng^{2}}{16(2\pi)^{2}}, \quad B \equiv \frac{Ng^{2}}{16(2\pi)^{2}}, \quad \lambda^{4} \equiv 2Ng^{2}\gamma^{4}.
\end{equation}
Solutions may be found in the form of power-laws:
\begin{equation} \label{anzatz1}
    G(q^{2}) = C_{G} (q^{2})^{-\alpha},
\end{equation}
\begin{equation} \label{anzatz2}
    \mathcal D(q^{2}) = C_{\mathcal D} (q^{2})^{-\beta},
\end{equation}
with $C_{G}$ and $C_{\mathcal D}$ -- dimensionless constants, $\alpha$ and $\beta$ -- some exponents.
It turned out, that straightforward substitution of anzatz (\ref{anzatz1}), (\ref{anzatz2}) in eqs. (\ref{Gh to deff}), (\ref{to diff}) suffers from certain mathematical difficulties and it is more convenient to do first some transformations in the equations above.
We divide equation (\ref{Gh to deff}) by $q^{2}$ then carry out differentiation with respect to $q^{2}$ in derived  equation and in eq. (\ref{to diff}), and obtain:
\begin{equation}
\frac{1}{q^{4}} \mathcal D^{-1}(q^{2}\!)\! +\! \frac{1}{q^{2}}\mathcal D^{-2}(q^{2})\mathcal D^{'}(q^{2})
\!\!= \!\!A  \mathcal D^{'}(q^{2}) \!\!\!\int_{0}^{q^{2}}\!\!\!\!\!\!\! dp^{2} p^{2} G(p^{2}).
\end{equation}\begin{equation}
    G^{-2}(q^{2}) G^{'}(q^{2})
    = \frac{\lambda^{4}}{q^{4}}
    - B \mathcal D^{'}(q^{2}) \!\!\int_{0}^{q^{2}}\!\!\!\!\!\!\! dp^{2}\, p^{4} \mathcal D(p^{2}).
\end{equation}
Now the anzatz (\ref{anzatz1}), (\ref{anzatz2}) can be applied. We substitute this power-laws into our equations and carry out all the integrations (under the assumption of $\beta \neq 3$ and $\alpha \neq 2$). Then we get
\begin{equation} \label{GSDE_alpha}
    (-\beta+1) C_{\mathcal D}^{-1} (q^{2})^{\beta-2}
    + \frac{\beta A C_{\mathcal D} C_{G}}{-\alpha+2} (q^{2})^{-\beta-\alpha+1}
    =0;
\end{equation}
\begin{equation}\label{GSDE_beta}
    - \alpha C_{G}^{-1} (q^{2})^{\alpha-1}
    - \lambda^{4} (q^{2})^{-2}
    - \frac{\beta B C_{\mathcal D}^{2}}{-\beta+3}  (q^{2})^{-2\beta+2}
    =0.
\end{equation}
We equate exponents of $q^{2}$ in both sides of this equations and obtain following equations for the parameters $\alpha$ and~ $\beta$:
\vspace{-.3cm}
\begin{equation}
    \beta-2 = -\beta-\alpha+1
\end{equation}
\begin{equation}
  \alpha-1 = -2 = -2\beta+2,
\end{equation}
which give us a unique solution
\begin{equation}
    \alpha = -1 \qquad \beta = 2.
\end{equation}
In terms of the infrared critical exponents
\begin{equation}
    \alpha_{G} = \alpha-1, \quad \alpha_{\mathcal D} = \beta-1,
\end{equation}
(see (\ref{DressedG}), (\ref{DressedD})), one has
\begin{equation}
    \alpha_{G} = -2, \quad \alpha_{\mathcal D} = 1.
\end{equation}
We now turn to the equations (\ref{GSDE_alpha}) and (\ref{GSDE_beta}) in order to obtain the coefficients $C_{G}$ and $C_{\mathcal D}.$ Thus we get
\begin{equation}
    - C_{\mathcal D}^{-1}   + \frac{2}{3} A C_{\mathcal D} C_{G} = 0;
\end{equation}
\begin{equation}
    C_{G}^{-1} - \lambda^{4} - 2 B C_{\mathcal D}^{2}   = 0.
\end{equation}
Along with the notations (\ref{A}) this gives
\begin{equation} \label{Coef.G}
    C_{G} = \frac{1}{\lambda^{4}}\left[1 + 3 B A^{-1}\right]
          = \frac{2}{\lambda^{4}};
\end{equation}
\begin{equation}\label{Coef.D}
     C_{\mathcal D} = \sqrt{\frac{3}{2} A^{-1} C_{G}^{-1}}
                        = 4\frac{\pi\lambda^{2}}{g\sqrt{N}},
\end{equation}
with $\lambda^{4} \equiv 2Ng^{2}\gamma^{4},$ where $\gamma$ is the Gribov parameter.

Consequently, the solution of the SDEs (\ref{Euclidean Ghost SDE}) and (\ref{Euclidean Gluon SDE}) is the following infrared behaviour of the gluon and ghost propagators, respectively:
\begin{equation} \label{IBGlP}
    G(q^{2}) = \frac{2}{\lambda^{4}} q^{2},
\end{equation}
\begin{equation}\label{IBGhP}
    \mathcal D(q^{2}) = 4\frac{\pi\lambda^{2}}{g\sqrt{N}} \frac{1}{q^{4}}.
\end{equation}
Thus one can see that the one-loop results are in good qualitative
agreement with the tree-level ones. The coefficients (\ref{Coef.G}),
(\ref{Coef.D}) obtained at this order, differ from the zeroth-order
ones (\ref{G(1)}), (\ref{mathcal D(1)}) by numerical multiplier and,
hopefully, are more precise.

As can be checked by numerical computation, the approximation
(\ref{weak angle approx.}) have no influence on the qualitative
behaviour of the propagators, but rather gives some additional
constant terms, unimportant for the asymptotic.

\section{Discussion and conclusions}\label{Concl.}

Formalism of the generalized effective action with implemented
restriction to the Gribov horizon is used to investigate the
infrared behaviour of the gluon and ghost propagators. The system of
the SDEs on the gluon propagator (\ref{Equation on Gluon
propagator}), the ghost propagator (\ref{Equation on Ghost
propagator}) and the Gribov parameter (\ref{Gorizon condition full})
is derived in this approach. These equations are obtained using
two-loop approximation for the generalized effective action.

The SDE for the Gribov parameter is in fact the full horizon
condition, which reduces to the Gribov's one at zeroth order.
Furthermore, the gluon propagator in this (zeroth-order)
approximation is exactly the same expression, first pointed out by
Gribov. The expression for the ghost propagator in zeroth-order
approximation agrees qualitatively with the Gribov's one but has
different numerical coefficient.

The obtained SDEs are also solved in the first-order approximation.
 The infrared critical exponents and coefficients for the gluon and ghost
propagators are obtained. Thus, we derive the expressions, which
describe the infrared behaviour of the gluon (\ref{IBGlP}) and ghost
(\ref{IBGhP}) propagators. The solution for the infrared exponents
was obtained before \cite{Zwanziger2}, whereas the coefficients for
the propagators at this order are derived first in the present work.

As expected, the first-order results agree with the zeroth-order
ones with different numerical coefficients at small momenta. For the
gluon propagator this difference reduces to $1/2$ and for the ghost
one - to $\sqrt{2}$.

Thus, we can see that cutting off the measure of the functional
integral at the (first) Gribov region deeply modifies the structure
of the propagators. This leads to infrared enhancement of the ghost
propagator and to suppression of the gluon one, as can be seen from
our calculations, described above. It is significant, that such
behaviour of the propagators supports the picture of color
confinement, \cite{Gribov1, Zwanziger, Zwanziger_conf2}, which we
believe to occur.

It would be interesting to study the possibility of generating the
gluon mass \cite{Dyn.Gen} in the effective action formalism of CJT
when the Zwanziger horizon condition is implemented \cite{Gr.Hor.}.

\section*{Acknowledgment}
I would like to thank Prof. V.P. Gusynin for the formulation of the problem and for the scientific support in all  stages of this work.

\end{document}